\documentclass[twocolumn,floatfix,showpacs,superscriptaddress]{revtex4}
\usepackage{graphicx}
\usepackage{amsmath}
\usepackage{bm}
\usepackage{color}
\begin{document}

\title{Extinction of coherent backscattering by a disordered photonic crystal with a Dirac spectrum}
\author{R. A. Sepkhanov}
\affiliation{Instituut-Lorentz, Universiteit Leiden, P.O. Box 9506, 2300 RA Leiden, The Netherlands}
\author{A. Ossipov}
\affiliation{School of Mathematical Sciences, 
University of Nottingham, 
University Park, 
Nottingham, NG7 2RD, United Kingdom}
\author{C. W. J. Beenakker}
\affiliation{Instituut-Lorentz, Universiteit Leiden, P.O. Box 9506, 2300 RA Leiden, The Netherlands}
\date{October 2008}
\begin{abstract}
Photonic crystals with a two-dimensional triangular lattice have a conical singularity in the spectrum. Close to this so-called Dirac point, Maxwell's equations reduce to the Dirac equation for an ultrarelativistic spin-$1/2$ particle. Here we show that the half-integer spin and the associated Berry phase remain observable in the presence of disorder in the crystal. While constructive interference of a scalar (spin-zero) wave produces a coherent backscattering \textit{peak}, consisting of a doubling of the disorder-averaged reflected photon flux, the destructive interference caused by the Berry phase \textit{suppresses} the reflected intensity at an angle which is related to the angle of incidence by time-reversal symmetry. We demonstrate this extinction of coherent backscattering by a numerical solution of Maxwell's equations and compare with analytical predictions from the Dirac equation.
\end{abstract}
\maketitle

Coherent backscattering is a rare example of an optical interference effect that is systematically constructive in a random medium \cite{Akk07}. A reciprocal pair of waves (related by time reversal symmetry) arrive in phase at the observer regardless of the path length or scattering sequence. By measuring the angular profile of the diffusively reflected intensity for plane wave illumination and averaging over the random scattering, a peak is observed at a specific reflection angle \cite{Kug84,Alb85,Wol85}. Under optimal conditions (same polarization of incident and reflected wave, transport mean free path $l$ much larger than wave length $\lambda$), the average peak and background intensity have ratio ${\cal R}=2$ \cite{Akk85,Mar88}. This factor of two enhancement follows directly by comparing the \textit{coherent} addition of intensities (first sum the two wave amplitudes, then square to obtain the total intensity) with the \textit{incoherent} addition (first square, then sum).

Because coherent backscattering does not depend on the random scattering phase shifts, it is a sensitive tool to probe systematic phase shifts that contain information about internal degrees of freedom of the scatterers and the photons. One example in light scattering from cold atoms is the coupling of the photon polarization to the collective spin of the atomic ensemble, which can change the constructive into a destructive interference \cite{Kup04}. This effect is similar to the change from ${\cal R}=2$ to ${\cal R}=1/2$ in electronic systems with strong spin-orbit coupling \cite{Ber88}.

A few years ago, Bliokh \cite{Bli05} discussed an altogether different and more dramatic switch from constructive to destructive interference: ultrarelativistic fermions (with an energy much greater than the rest energy) would have a complete extinction of coherent backscattering, so ${\cal R}=0$, as a consequence of the Berry phase of $\pi$ accumulated along a closed trajectory by a half-integer spin pointing in the direction of motion. Indeed, it was previously noticed by Ando et al.\ in the context of graphene \cite{And98} that the scattering amplitude $s(\phi)$ for massless Dirac fermions vanishes when $\phi\rightarrow\pi$ (with $\phi$ the angle between the initial and final wave vectors $\bm{k}_{i}$ and $\bm{k}_{f}$). This absence of backscattering can be understood either in terms of the Berry phase difference of $\pi$ between time reversed scattering sequences \cite{And98} or in terms of the antisymmetry $S(\bm{k}_{i}\rightarrow\bm{k}_{f})=-S(-\bm{k}_{f}\rightarrow -\bm{k}_{i})$ of the scattering matrix of the Dirac equation \cite{Suz02}. Absence of backscattering in graphene might be measurable if scanning probe microscopy can provide the required angular resolution of the electron flow \cite{Chi08}.

Here we investigate an alternative realization of the extinction of coherent backscattering that relies on photons rather than ultrarelativistic fermions. The half-integer spin required for the Berry phase of $\pi$ is produced by the Dirac-type band structure of a triangular-lattice photonic crystal \cite{Hal08,Sep08}. By using photons rather than electrons the difficulty of angular resolved detection is avoided. We first discuss the effect at the level of the Dirac equation, and then test the predictions with a numerical solution of the full Maxwell's equations.

\begin{figure}[tb]
\centerline{\includegraphics[width=0.6\linewidth]{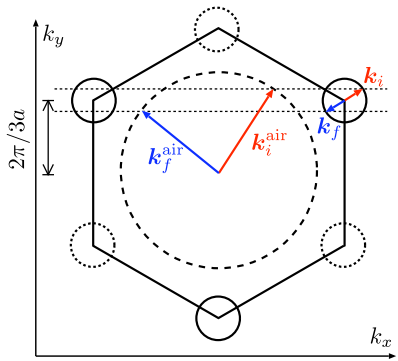}}
\caption{\label{BZ}
Hexagonal first Brillouin zone of a two-dimensional triangular lattice photonic crystal, with equifrequency contours centered at the corners. The three solid circles are related by translation over a reciprocal lattice vector, so they are equivalent, while scattering from a solid to a dotted circle is suppressed if the scattering potential is smooth on the scale of the lattice constant $a$. The large dashed circle at the center is the equifrequency contour in air, included to indicate the refraction at the air-crystal interface (as expressed by Eq.\ \eqref{krelation}).
}
\end{figure}

We consider a photonic crystal with a two-dimensional (2D) triangular lattice structure. The hexagonal first Brillouin zone is shown in Fig.\ \ref{BZ}. Haldane and Raghu \cite{Hal08} showed that a pair of almost degenerate envelope Bloch waves $(\Psi_{1},\Psi_{2})\equiv\Psi$ near a corner of the Brillouin zone can be represented by a pseudospin, coupled to the orbital motion. On length scales large compared to the lattice constant $a$ and for frequencies near the degeneracy frequency $\omega_{D}$, the wave equation reduces to
\begin{equation}
\left(-i\sigma_{x}\frac{\partial}{\partial x}-i\sigma_{y}\frac{\partial}{\partial y}\right)\Psi=\frac{\omega-\omega_{D}}{v_{D}}\Psi,\label{HDirac}
\end{equation}
with Pauli matrices $\sigma_{x},\sigma_{y}$. This is the 2D Dirac equation of a spin-$\tfrac{1}{2}$ particle with zero mass and group velocity $v_{D}$ (of order $a\omega_{D}$). The dispersion relation
\begin{equation}
(\omega-\omega_{D})^{2}=v_{D}^{2}(k_{x}^{2}+k_{y}^{2})\label{Diracdispersion}
\end{equation}
has a double cone with a degeneracy at frequency $\omega_{D}$ (the so-called Dirac point).

The Berry phase associated with the pseudospin degree of freedom was calculated in Ref.\ \cite{Sep08}.  The solution of Eq.\ \eqref{HDirac} with a definite wave vector $\bm{k}=(k_{x},k_{y})$ is
\begin{equation}
\Psi={\cal C}^{-1/2}\begin{pmatrix}
(\omega-\omega_{D})/v_{D}\\
k_{x}+ik_{y}
\end{pmatrix}
\equiv 
\begin{pmatrix}
\cos(\theta/2)\\
e^{i\phi}\sin(\theta/2)
\end{pmatrix},\label{PsiBloch}
\end{equation}
with $\cal{C}$ a normalization constant. The angles $\phi,\theta$ define the Bloch vector $(\cos\phi\sin\theta,\sin\phi\sin\theta,\cos\theta)$, representing the direction of the pseudospin on the Bloch sphere. Because of the dispersion relation \eqref{Diracdispersion} the angle $\theta=\pi/2$, so the Bloch vector lies in the $x-y$ plane, pointing in the direction of $\bm{k}$. The Berry phase $\phi_{B}$ is one half the solid angle subtended at the origin by the rotating Bloch vector. A rotation of $\bm{k}$ by $360^{\circ}$ in the $x-y$ plane thus produces a Berry phase $\phi_{B}=\pi$.

Because of refraction at the interfaces $x=0$ and $x=L$ between the photonic crystal and air, we need to distinguish the initial and final wave vectors $\bm{k}_{i}$, $\bm{k}_{f}$ of envelope Bloch waves inside the crystal (velocity $v_{D}$) from the corresponding values $\bm{k}_{i}^{\rm air},\bm{k}_{f}^{\rm air}$ of plane waves outside (velocity $c$). For the crystallographic orientation shown in Fig. \ref{BZ}, the wave vectors before and after refraction (at a given frequency $\omega$) are related by \cite{Sep07}
\begin{subequations}
\label{krelation}
\begin{align}
&k_{y}^{\rm air}=k_{y}+2\pi/3a,\label{kyrelation}\\
&c^{2}(k^{\rm air})^{2}-\omega^{2}=0=v_{D}^{2}k^{2}-(\omega-\omega_{D})^{2},\label{kxrelation}
\end{align}
\end{subequations}
with $a$ the lattice constant and $\omega_{D}$ the frequency of the Dirac point.
We denote $k^{\rm air}=|\bm{k}^{\rm air}|$ and $k=|\bm{k}|$.

Before considering the diffuse reflection from the disordered photonic crystal, we address the specular reflection from the air-crystal interface that is present even without any disorder. A plane wave incident at an angle
\begin{equation}
\theta=\arcsin(ck_{i,y}^{\rm air}/\omega)=\arcsin(2\pi c/3\omega a+ck_{i,y}/\omega)\label{thetakyrelation}
\end{equation}
is specularly reflected at an angle $\theta_{\rm spec}=\pi-\theta$. The interface reflectivity $R_{\rm int}$ (the fraction of the incident photon flux that is specularly reflected) follows from the transfer matrix of the air-crystal interface calculated in Ref.\ \cite{Sep07}. In the approximation of maximal coupling we find
\begin{equation}
R_{\rm int}(k_{y})=\frac{1-\sqrt{1-(k_{y}/k)^{2}}}{1+\sqrt{1-(k_{y}/k)^{2}}},\label{Rresult}
\end{equation}
hence the interface reflectivity is zero for $k_{y}=0$. Numerical solutions of Maxwell's equations \cite{Sep07} give $R_{\rm int}(0)\simeq 0.05$ for $\omega$ near $\omega_{D}$, so this is a reasonably accurate approximation.   

Disorder inside the photonic crystal produces a background of diffusively reflected waves in an angular opening $\delta\theta\simeq k/k^{\rm air}\ll 1$ around $\theta_{0}=\pi-\arcsin(2\pi c/3\omega a)$. The reciprocity angle $\theta_{\ast}$ for coherent backscattering is related to the incident angle $\theta$ of Eq.\ \eqref{thetakyrelation} by
\begin{equation}
\theta_{\ast}=\pi-\arcsin(2\pi c/3\omega a-ck_{i,y}/\omega).\label{thetastarkyrelation}
\end{equation}
We will choose $k_{i,y}$ small ($\ll k$) but nonzero, so that $R_{\rm int}\ll 1$ while still coherent backscattering at angle $\theta_{\ast}$ can be resolved from specular reflection at angle $\theta_{\rm spec}$.

\begin{figure}[tb]
\centerline{\includegraphics[width=0.6\linewidth]{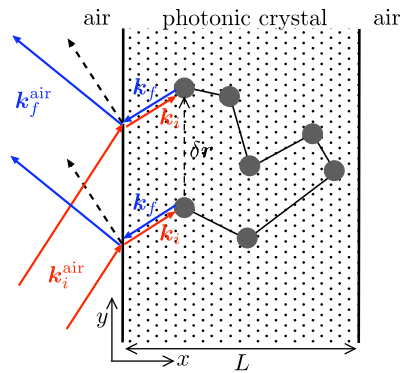}}
\caption{\label{trajectories}
A pair of reciprocal waves that interfere destructively, resulting in the extinction of coherent backscattering. Arrows at the air-crystal interface indicate the incident plane wave (red solid lines), the diffusively reflected wave (blue solid lines), and the specularly reflected wave (black dashed lines). The two initial waves (wave vector $\bm{k}_{i}^{\rm air}$, refracted to $\bm{k}_{i}$) follow time-reversed sequences of scattering events (dark circles), to end up in two final waves (wave vector $\bm{k}_{f}$, refracted to $\bm{k}_{f}^{\rm air}$) with a phase difference of $\pi+(\bm{k}_{i}+\bm{k}_{f})\cdot\delta\bm{r}$. For $\bm{k}_{f}=-\bm{k}_{i}$ only the Berry phase difference of $\pi$ remains. As a consequence of this destructive interference, the intensity of the reflected wave indicated in blue is suppressed. By measuring the angular profile of the average reflected intensity a minimum of nearly zero intensity will result at this angle.
}
\end{figure}

The extinction of coherent backscattering by destructive interference of reciprocal waves is illustrated in Fig.\ \ref{trajectories}. The two series of time reversed scattering events $S_{+}=\bm{k}_{i}\rightarrow\bm{k}_{1}\rightarrow\bm{k}_{2}\rightarrow\cdots\rightarrow\bm{k}_{n-1}\rightarrow\bm{k}_{n}\rightarrow-\bm{k}_{i}$ and $S_{-}=\bm{k}_{i}\rightarrow-\bm{k}_{n}\rightarrow-\bm{k}_{n-1}\rightarrow\cdots\rightarrow-\bm{k}_{2}\rightarrow-\bm{k}_{1}\rightarrow-\bm{k}_{i}$ have the same scattering amplitude up to a Berry phase difference $\phi_{B}=\pi$ \cite{Bli05,And98}. This destructive interference suppresses the reflected intensity at angle $\theta_{\ast}$. If the final wave vector $\bm{k}_{f}$ deviates from the exact backscattering direction $-\bm{k}_{i}$, the phase difference $\Delta\phi=\phi_{B}+(\bm{k}_{i}+\bm{k}_{f})\cdot\delta\bm{r}$ depends on the separation $\delta\bm{r}$ of the first and last scattering events \cite{Akk85}. The Berry phase difference $\phi_{B}$ remains equal to $\pi$, because both the Berry phases accumulated along $S_{+}$ and $S_{-}$ are incremented by the same amount (half the angle between $\bm{k}_{f}$ and $-\bm{k}_{i}$). 

By including the Berry phase in the theory \cite{Akk07,Akk85,Mar88} of coherent backscattering of scalar waves in the weak scattering regime ($l\gg\lambda$), it follows that the incoherent part $R_{0}$ of the reflectivity $R$ remains unaffected while the interference part $\delta R$ acquires an overall factor $\cos\phi_{B}$ \cite{Bli05}:
\begin{equation}
R=R_{0}+\delta R\cos\phi_{B}.\label{RphiB}
\end{equation}
The entire angular profile of coherent backscattering in the Dirac equation (where $\phi_{B}=\pi$) may therefore be obtained, for $l\gg\lambda$, by simply changing the sign of the known results for $\delta R$ for scalar waves (where $\phi_{B}=0$). In particular, for $k_{i,y}\ll k$ and $\delta k=k_{f,y}+k_{i,y}\ll 1/l$ one has
\begin{align}
R_{0}={}&\frac{1}{\pi N}(1+z_{0}/l)\left(1-\frac{l+z_{0}}{L+2z_{0}}\right),\label{R0result}\\
\delta R=&{}\frac{1}{\pi N}(1+z_{0}/l)\nonumber\\
&\times\bigl\{1-(l+z_{0})\delta k\coth[\delta k(L+2z_{0})]\bigr\}.\label{deltaRresult}
\end{align}
The normalization factor $N=kW/2\pi$ is chosen such that $R$ is the fraction of the incident photon flux which is reflected in a single transverse mode when $k_{f,y}=2\pi n/W$ $(n=0,\pm 1, \pm 2,\ldots)$ is discretized by periodic boundary conditions at $y=0$ and $y=W$. (This normalization is chosen to simplify the comparison with the numerical calculations described later on.)

Eqs.\ \eqref{R0result} and \eqref{deltaRresult} are approximate results from the radiative transfer equation, accurate for a disordered slab of thickness $L$ not much smaller than the transport mean free path $l$. The parameter $z_{0}$, the so-called extrapolation length, depends on the reflectivity $R_{\rm int}$ of the interface between the photonic crystal and vacuum, according to \cite{Zhu91}
\begin{subequations}
\label{z0result}
\begin{align}
&z_{0}=\tfrac{1}{4}\pi l\,\frac{1+C_{2}}{1-C_{1}},\;\;
C_{1}=\frac{1}{k}\int_{0}^{k}R_{\rm int}(k_{y})dk_{y},\label{C1def}\\
&C_{2}=\frac{4}{\pi}\int_{0}^{k}\sqrt{1-(k_{y}/k)^{2}}\,R_{\rm int}(k_{y})dk_{y}.\label{C2def}
\end{align}
\end{subequations}
Substitution of Eq.\ \eqref{Rresult} gives the simple answer $z_{0}=l$ for the extrapolation length of the air-crystal interface.

Collecting results, we arrive at the following line shape of the reflectivity near the reciprocity angle:
\begin{align}
R&=\frac{4}{\pi N}\left(l\delta k\coth[\delta k(L+2l)]-\frac{l}{L+2l}\right)\nonumber\\
&\rightarrow \frac{4}{3\pi N}(Ll+2l^{2})\delta k^{2}\;\;{\rm for}\;\;\delta k\rightarrow 0.\label{Rfinal}
\end{align}

The suppression \eqref{Rfinal} of the reflected intensity within a narrow angular opening $\delta\theta\simeq 1/kl$ around the reciprocity angle $\theta_{\ast}$ is compensated by an excess intensity $\delta R\simeq 1/kl$ of the diffusively reflected wave at angles away from $\theta_{\ast}$. This compensation is required by current conservation \cite{Fie08}, but difficult to observe for $kl\gg 1$. We will therefore not consider it in what follows.

We now compare the analytical predictions from the Dirac equation with a numerical solution of Maxwell's equations. As in earlier work \cite{Sep07}, we use the {\sc meep} software package \cite{Far06} to solve Maxwell's equations in the time domain \cite{Taf05} for a continuous plane wave source (time dependence $\propto e^{i\omega t}$) switched on gradually. We calculate the reflected intensity, projected onto transverse modes, as a function of time and take the large time limit to obtain the stationary reflectivity $R$. We took the time sufficiently large that the sum of the total transmission and total reflection differs from unity by less than $0.03$.

The photonic crystal consists of a triangular lattice of parallel dielectric rods (dielectric constant $8.9$, radius $\rho_{0}=0.3\,a$) in air. The orientation of the lattice is as shown in Fig.\ \ref{trajectories}. The magnetic field is taken parallel to the rods (TE polarization). The conical singularity in the band structure is at frequency $\omega_D=3.03\,c/a$, with a slope  $d\omega/dk\equiv v_D=0.432\,c$. We discretize the transverse wave vector by means of periodic boundary conditions: $k_{y}=2\pi n/W$, $n=0,\pm 1,\pm 2,\ldots$, with $W=1200\,a$. The longitudinal dimension of the lattice is taken at $L=60\sqrt{3}\,a$ (corresponding to 121 rows of dielectric rods). The frequency $\omega$ of the incident plane wave is chosen such that $\omega-\omega_{D}=0.206\,c/a$ is sufficiently large that $N=88\gg 1$, but sufficiently small that the trigonal distortion of the circular equifrequency contours is insignificant. 

Disorder is introduced by randomly varying the radius $\rho(\bm{r})$ of the dielectric rods at position $\bm{r}$, according to $\rho(\bm{r})=\rho_{0}+\delta\rho(\bm{r})$. The variation $\delta\rho(\bm{r})$ is spatially correlated over a length $\xi$ in a Gaussian manner,
\begin{equation}
\delta\rho(\bm{r})=\sum_{i=1}^{\cal N}A_{i}\exp(-|\bm{r}-\bm{r}_{i}|^{2}/2\xi^{2}),\label{deltarho}
\end{equation}
where $\bm{r}_{i}$ ($i=1,2,\ldots{\cal N}$) is a randomly chosen point in the crystal and $A_{i}$ is a randomly chosen amplitude in the interval $(-\Delta,\Delta)$. We took ${\cal N}=3200$, $\Delta=0.04\,a$, $\xi=2a$. It is essential that the correlation length $\xi$ of the disorder is larger than the lattice constant $a$ to minimize scattering between the two inequivalent corners of the Brillouin zone (solid and dashed circles in Fig.\ \ref{BZ}). For these disorder parameters we found that only about $2\%$ of the incident flux is reflected into the opposite corner.

\begin{figure}[tb]
\centerline{\includegraphics[width=0.90\linewidth]{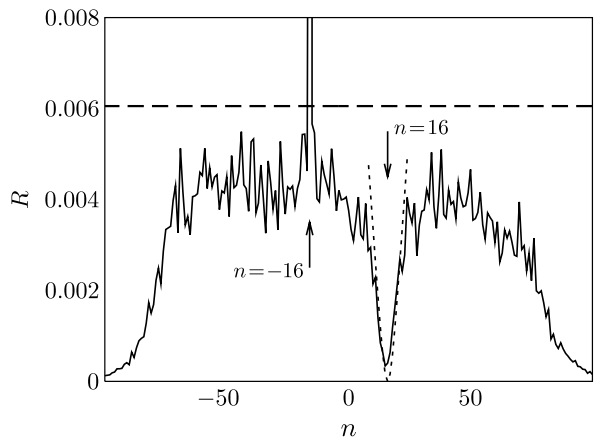}}
\caption{\label{fig_data}
Plot of the reflectivity $R$ (fraction of incident flux reflected in a single transverse mode) as a function of the transverse mode index $n$ (related to the transverse wave vector by $k_{f,y}=2\pi n/W$). The specularly reflected flux is at $n=-16$ and the extinction of coherent backscattering is at $n=+16$ (indicated by arrows). The curve is calculated numerically from Maxwell's equations. The dashed and dotted lines are, respectively, the analytical predictions \eqref{R0result} and \eqref{Rfinal} for the incoherent background and the line shape near the extinction angle, for a transport mean free path of $l=10\,a$.
}
\end{figure}

The numerical data in Fig.\ \ref{fig_data} is for the incident wave vector $k_{i,y}=-32\,\pi/W$, averaged over 80 realizations of the disorder. The angle of incidence (measured relative to the positive $x$-axis) is $\theta=0.67\,\mbox{rad}=38.4^{\circ}$. The specularly reflected wave at an angle $\theta_{\rm spec}=\pi-\theta=2.47\,\mbox{rad}=141.5^{\circ}$ has transverse mode number $n=-16$. The reciprocity angle \eqref{thetastarkyrelation} is $\theta^{\ast}=2.40\,\mbox{rad}=137.5^{\circ}$, corresponding to the mode number $n=+16$.

Both the specularly reflected wave at $n=-16$ and the extinction of the coherent backscattering at $n=+16$ are clearly visible in Fig.\ \ref{fig_data}. The extinction at the reciprocity angle is not complete, presumably because of scattering between inequivalent corners of the Brillouin zone. The analytical theory predicts a parabolic line shape of the reflectivity around the reciprocity angle, given by Eq.\ \eqref{thetastarkyrelation}, the width of which depends on the value of the transport mean free path $l$. A fit to the numerical data gives $l=10\,a$ (dotted curve in Fig.\ \ref{fig_data}), which then implies a background of incoherent diffuse reflection at a value $R_{0}\approx 0.006$ which is somewhat larger than the numerical data (dashed horizontal line). 

In conclusion, we have shown that coherent backscattering of radiation from a disordered triangular lattice photonic crystal is extinguished at an angle that is reciprocal to the angle of incidence. This effect has an analogue for electrons \cite{Bli05,And98}, but it should be easier to observe for photons because angular resolved detection is more feasible. The observation of the extinction of coherent backscattering would be a striking demonstration of a spin-$1/2$ Berry phase in a disordered optical system.

We acknowledge discussions with J. H. Bardarson and C. W. Groth. This research was supported by the Dutch Science Foundation NWO/FOM.

\end{document}